\documentclass[aps,groupedaddress,fleqn,twocolumn,longbibliography]{revtex4-2}
\usepackage{graphicx}
\usepackage{xcolor}
\usepackage{amsmath,amssymb}
\usepackage{float}
\usepackage{amsfonts}
\usepackage{verbatim}
\usepackage{flushend}
\usepackage{balance}
\usepackage{endnotes}
\usepackage{footnote}
\usepackage{adjustbox}
\usepackage{gensymb}
\usepackage{float}
\usepackage{epigraph}
\usepackage{xcolor}
\usepackage[utf8]{inputenc}
\usepackage{setspace}
\usepackage[colorlinks=true,citecolor=blue,urlcolor=blue]{hyperref}

\newcommand{\beq}{\begin{eqnarray}}
\newcommand{\eeq}{\end{eqnarray}}
\usepackage{amsmath}

\begin{document}

\author{K. Trachenko}
\thanks{k.trachenko@qmul.ac.uk}
\affiliation{School of Physical and Chemical Sciences, Queen Mary University of London, 327 Mile End Road, London, E1 4NS, United Kingdom}
\author{B. Monserrat}
 \affiliation{Department of Materials Science and Metallurgy, University of Cambridge, 27 Charles Babbage Road, Cambridge CB3 0FS, United Kingdom}
 \affiliation{Cavendish Laboratory, University of Cambridge, J. J. Thomson Avenue, Cambridge CB3 0HE, United Kingdom}
\author{M. Hutcheon}
 \affiliation{Intellectual Ventures, Bellevue, Washington, United States}
 \author{Chris J. Pickard}
 \affiliation{Department of Materials Science and Metallurgy, University of Cambridge, 27 Charles Babbage Road, Cambridge CB3 0FS, United Kingdom}
\affiliation{Advanced Institute for Materials Research, Tohoku University, Sendai, Japan}
\title{Upper bounds on the highest phonon frequency and superconducting temperature from fundamental physical constants}

\begin{abstract}
Fundamental physical constants govern key effects in high-energy particle physics and astrophysics, including the stability of particles, nuclear reactions, formation and evolution of stars,
synthesis of heavy nuclei and emergence of stable molecular structures. Here, we show that fundamental constants also set an upper bound for the frequency of phonons in condensed matter phases, or
how rapidly an atom can vibrate in these phases. This bound is in agreement with \textit{ab initio} simulations of atomic hydrogen and high-temperature hydride superconductors, and implies an upper
limit to the superconducting transition temperature $T_c$ in condensed matter. Fundamental constants set this limit to the order of 10$^2-10^3$ K. This range is consistent with our calculations of
$T_c$ from optimal Eliashberg functions. As a corollary, we observe that the very existence of the current research of finding $T_{\mathrm{c}}$ at and above $300$ K is due to the observed values of
fundamental constants. We finally discuss how fundamental constants affect the observability and operation of other effects and phenomena including phase transitions.
\end{abstract}

\maketitle

\section{Introduction}

The governing role of fundamental physical constants has been extensively discussed in high-energy particle physics and astrophysics. Examples include the stability of nuclei, nuclear reactions, and the formation and evolution of stars -- where heavy elements are produced and give rise to other observable structures. In these and other processes, fundamental constants give the Universe its observed properties and separate it from potential other universes where these constants may be different \cite{barrow,barrow1,carr,carrbook,finebook,cahnreview,hoganreview,adamsreview,uzanreview}. Understanding the origin of fundamental constants and their fine-tuning is considered one of the grand challenges in science\,\cite{grandest,weinberg}. Approaching this challenge starts with deepening our understanding of how fundamental constants affect phenomena spanning the length and energy scales between particle physics and astrophysics\,\cite{advphysreview,brareview,myropp}.

In condensed matter physics and in other areas that support a continuum description of matter, such as elasticity and hydrodynamics, system properties are driven by many-body collective effects.
These effects are often considered as emergent and thus not reducible to individual particles whose properties are set by fundamental constants. It was therefore unexpected when it was discovered
that fundamental constants govern the {\it bounds}, characteristic values and properties of several key many-body collective processes. Examples include viscosity\,\cite{kss,sciadv,zaccone-visc},
diffusion and spin dynamics\,\cite{spin,hbarm1,hbarm2,hbarm3,hbarm4}, electron and heat transport\,\cite{hartnoll,hartnoll1,behnia,nussinov1}, thermal conductivity in insulators\,\cite{momentumprb}
and conductors\,\cite{brareview}, the speed of sound\,\cite{sciadv2}, elastic moduli including those in lower-dimensional systems\,\cite{brareview,advphysreview}, thermal expansion,
melting\,\cite{weisalpha,premelting} and velocity gradients that can be set up in cells using biochemical energy\,\cite{sciadv2023}. In these examples, properties are governed by either fundamental
constants only or by a combination of these constants with other parameters such as temperature. The implications of these findings extend to both fundamental and applied
science\,\cite{advphysreview}.

Here, we show that fundamental physical constants provide an upper bound to the frequency of phonons in condensed matter phases, setting a limit to the vibrational frequency of atoms in these phases.
This bound involves the dimensionless electron-to-proton mass ratio $\beta$, the key constant involved in setting the finely-tuned ``habitable zone'' of our world in the ($\alpha$, $\beta$) space,
where $\alpha$ is the fine-structure constant \cite{barrow}. The bound applies to non-molecular systems, and we find it to be in agreement with \textit{ab initio} simulations of atomic hydrogen and
hydride superconductors which, as hydrogen is the lightest nucleus (a single proton), host the highest possible phonon frequencies. We subsequently show that this upper bound in the maximal phonon
frequency implies an upper bound to the phonon-mediated superconducting critical temperature $T_{\mathrm{c}}$ in condensed matter. In particular, we show that fundamental constants set the upper
bound to $T_c$ on the order of $10^2-10^3$\,K, consistent with calculations of $T_c$ from optimal Eliashberg functions. We also observe that in addition to electron-phonon coupling, the bound to
$T_c$ in terms of fundamental constants applies to other coupling mechanisms such as electron-electron coupling. This implies that the current search for $T_{\mathrm{c}}$ at and above $300$\,K is
itself due to fundamental constants taking their observed values. We finally observe that fundamental constants affect the observability and operation of other phenomena in condensed matter physics
including phase transitions. This has wider implications compared to earlier work discussing how fundamental constants set bounds on properties that are already observed
\cite{kss,sciadv,spin,hbarm1,hbarm2,hbarm3,hbarm4,hartnoll,hartnoll1,behnia,nussinov1,momentumprb,sciadv2,brareview,sciadv2023,advphysreview,myropp}.

\section{Methods}

In our density functional theory calculations, we use similar methods to our earlier paper \cite{sciadv2}. This includes using the Perdew-Burke-Ernzerhof (PBE) exchange-correlation functional
\cite{pbe}, an energy cutoff of $1200$\,eV and a $\mathbf{k}$-point grid of spacing $2\pi\times0.025$\,\AA$^{-1}$ to sample the electronic Brillouin zone.

We relax cell parameters and internal coordinates to obtain a pressure to within $10^{-4}$ GPa of the target pressure and forces smaller than $10^{-5}$\,eV/\AA. We calculate the phonon dispersion
using the finite displacement method \cite{fd_martin} in conjunction with nondiagonal supercells \cite{nondiagonal} with a $4\times4\times4$ coarse $\mathbf{q}$-point grid to sample the vibrational
Brillouin zone. We use Fourier interpolation to calculate the phonon frequencies along a high symmetry path in the Brillouin zone.

\section{Upper bound on the vibrational frequency}

We start with the upper bound on the vibrational frequency. Oscillations are ubiquitous in nature and play a central role in all areas of physics, including condensed matter physics. The question of how high the oscillation frequency of an atom can be in solids or liquids is therefore of general importance.

The maximal frequency in a given condensed system, $\omega_{\mathrm{m}}$, is the largest frequency in the phonon dispersion, and is typically close to the Debye frequency $\omega_{\rm D}$. We are
interested in an upper bound, $\omega_{\mathrm{m}}^{\mathrm{u}}$, on this maximal frequency (and thus an upper bound on \textit{all} phonon frequencies). We recall two important properties of
condensed matter phases: the interatomic separation $a$ and a cohesive, or bonding, energy $E$. The Bohr radius, $a_{\rm B}$, sets the characteristic scale of $a$ to the order of Angstroms:

\begin{equation}
a_{\rm B}=\frac{4\pi\epsilon_0\hbar^2}{m_e e^2}
\label{bohr}
\end{equation}

\noindent and the Rydberg energy, $E_{\rm R}=\frac{e^2}{8\pi\epsilon_0a_{\rm B}}$ \cite{ashcroft}, sets the characteristic scale of $E$ to the order of several eV:

\begin{equation}
E_{\rm R}=\frac{m_ee^4}{32\pi^2\epsilon_0^2\hbar^2}
\label{rydberg}
\end{equation}

\noindent where $e$ and $m_e$ are electron charge and mass.

The ratio $\frac{\hbar\omega_{\mathrm m}}{E}$ can be derived by approximating $\hbar\omega_{\mathrm m}$ as $\hbar\left(\frac{E}{ma^2}\right)^{\frac{1}{2}}$, where $m$ is the atom mass and then either
(a) using $a=a_{\rm B}$ from (\ref{bohr}) and $E=E_{\rm R}$ from (\ref{rydberg})  or (b) using $E=\frac{\hbar^2}{2m_ea^2}$. This gives, up to a factor of order one ($\sqrt{2}$), the following
relation:

\begin{equation}
\frac{\hbar\omega_{\rm m}}{E}=\left(\frac{m_e}{m}\right)^{\frac{1}{2}}
\label{ratio}
\end{equation}

The same ratio \eqref{ratio} follows by combining two known relations in metallic systems: $\frac{\hbar\omega_{\rm m}}{E}\approx\frac{c}{v_{\rm F}}$, where $v_{\rm F}$ is the Fermi velocity and $c$ is the speed of sound, and $\frac{c}{v_{\rm F}}\approx\left(\frac{m_e}{m}\right)^{\frac{1}{2}}$, providing an order-of-magnitude estimation of $\frac{\hbar\omega_{\rm m}}{E}$ in other systems too \cite{ashcroft}.

The upper bound $\omega_{\rm m}^{\rm u}$ is obtained from  Eq.\,\eqref{ratio} by setting $m=m_p$, the proton mass, corresponding to atomic hydrogen. Combining this with Eq. \eqref{rydberg} yields finally

\begin{equation}
\omega_{\rm m}^{\rm u}=\frac{1}{32\pi^2\epsilon_0^2}\frac{m_ee^4}{\hbar^3}\left(\frac{m_e}{m_p}\right)^{\frac{1}{2}}
\label{ratio2}
\end{equation}

Interestingly, $\omega_{\rm m}^{\rm u}$ in \eqref{ratio2} can also be written in terms of the Compton frequency $\omega_{\rm C}=\frac{m_ec^2}{\hbar}$ and the fine structure constant
$\alpha=\frac{1}{4\pi\epsilon_0}\frac{e^2}{\hbar c}$ as

\begin{equation}
\omega_{\rm m}^{\rm u}=\frac{1}{2}\omega_{\rm C}\alpha^2\left(\frac{m_e}{m_p}\right)^{\frac{1}{2}}
\label{ratio3}
\end{equation}

Equation\,\,\eqref{ratio3} offers a simple result showing that the upper bound to the atomic oscillation frequency is set by the Compton frequency times the square of $\alpha$ and the square-root of the electron-to-proton mass ratio.

Equation\,\,\eqref{ratio2} sets $\omega_{\rm m}^{\rm u}$ in terms of fundamental physical constants. The constants $\hbar$, $m_e$, and $e$ determine the electromagnetic energy scale driving atomic oscillations. The proton mass $m_p$ is intrinsic to the atoms being driven and sets the scale (both in time, and space) of the oscillatory response. Interestingly, $\omega_{\rm m}^{\rm u}$ involves the electron-to-proton mass ratio $\beta=\frac{m_e}{m_p}$. This dimensionless constant is considered of fundamental importance: together with the fine-structure constant, it sets a finely-tuned habitable zone in ($\alpha$, $\beta$) space where it governs the stability of molecular structures, processes involved in igniting and fuelling stars (and thus the production of heavy elements), and other essential effects giving rise to our observable world\,\cite{barrow}.

With $E_{\rm R}$=13.6\,eV and $\beta=\frac{m_e}{m_p}=\frac{1}{1836}$, $\omega_{\rm m}^{\rm u}$ in Eq.\,\eqref{ratio2} is about (here and below, $k_{\rm B}=1$):
\begin{equation}
\omega_{\rm m}^{\rm u}=3680\,\rm {K}
\label{omegal}
\end{equation}

The upper bound $\omega_{\rm m}^{\rm u}$ in \eqref{omegal} is of the same order of magnitude, but above, the experimental Debye frequency $\omega_{\rm D}$ of 2240\,K in ambient pressure diamond
\cite{diamonddebye}, the hardest known material\,\cite{diamondvadim} or a similar value of the maximal phonon frequency in graphene \cite{graphene-omega}.

The upper bound \eqref{ratio2}-\eqref{omegal} corresponds to solid hydrogen with strong metallic bonding. Although this phase only exists at megabar pressures\,\cite{silvera,loubeyre,hydrogen} and is
thermodynamically and dynamically unstable at ambient pressure, it is interesting to check the validity of our bound and calculate $\omega_{\rm m}$ in atomic hydrogen. The importance of this
calculation is also highlighted by the strong interest in the properties of atomic hydrogen at high pressure (see, e.g., Refs. \cite{silvera,loubeyre,hydrogen}).

Before calculating $\omega_{\rm m}$, we first evaluate the effect of pressure on the fundamental bound \eqref{ratio2} because solid atomic hydrogen exists at high pressure only as mentioned above. We
make two remarks in relation to the pressure effect. First, hydrogen is a unique element with no core electrons. As compared to heavier elements, this gives weaker repulsive contributions to the
interatomic interaction and weaker pressure dependence of elastic moduli\,\cite{vadim2}. As a result, pressure has a weaker effect on $\omega_{\rm m}$ in hydrogen as compared to other systems.
Second, the effect of pressure on $\omega_{\rm m}^{\rm u}$ can be approximately estimated by adding the term $PV$ to $E$ in Eq.\,\eqref{ratio}, representing the work needed to overcome the elastic
deformation\,\cite{frenkel} due to external pressure in order to surmount the energy of cohesion. Here, $V$ is the elementary volume on the order of $a^3$. According to Eq. \eqref{ratio}, this
increases $\omega_m^u(P)$ as

\begin{equation}
\omega_{\rm m}^{\rm u}(P)=\omega_{\rm m}^{\rm u}+\left(\frac{m_e}{m_p}\right)^{\frac{1}{2}}\frac{a^3P}{\hbar}
\label{externalp}
\end{equation}

We note that the full treatment of pressure dependence of frequency for a particular system involves the system-specific equation of state and may involve nonlinear elasticity
\cite{zaccone-pressure}. This introduces system-specificity in $\omega_{\rm m}^{\rm u}(P)$. Our aim here is to evaluate the effect of pressure in terms of fundamental constants only. This evaluation
is necessarily approximate and provides an order-of-magnitude estimation \cite{advphysreview,brareview}. We will later see that setting $a$ close to $a_{\rm B}$ agrees with the modelling results,
showing that our approximate treatment is sensible.

We calculate the phonon dispersion curves of atomic hydrogen in the $I4_1/amd$ structure \cite{i41amd,pickard_h_natphys}, which is currently the best candidate structure for solid atomic metallic
hydrogen. This structure is calculated to become thermodynamically stable in the pressure range $400$--$600$ GPa \cite{azadi_metal,morales_metal,mauri-stability}, below which solid hydrogen is a
molecular solid. We perform density functional theory (DFT) calculations using the {\sc castep} package \cite{castep}. Details of DFT calculations are given in the Appendix.

\begin{figure}
{\scalebox{0.33}{\includegraphics{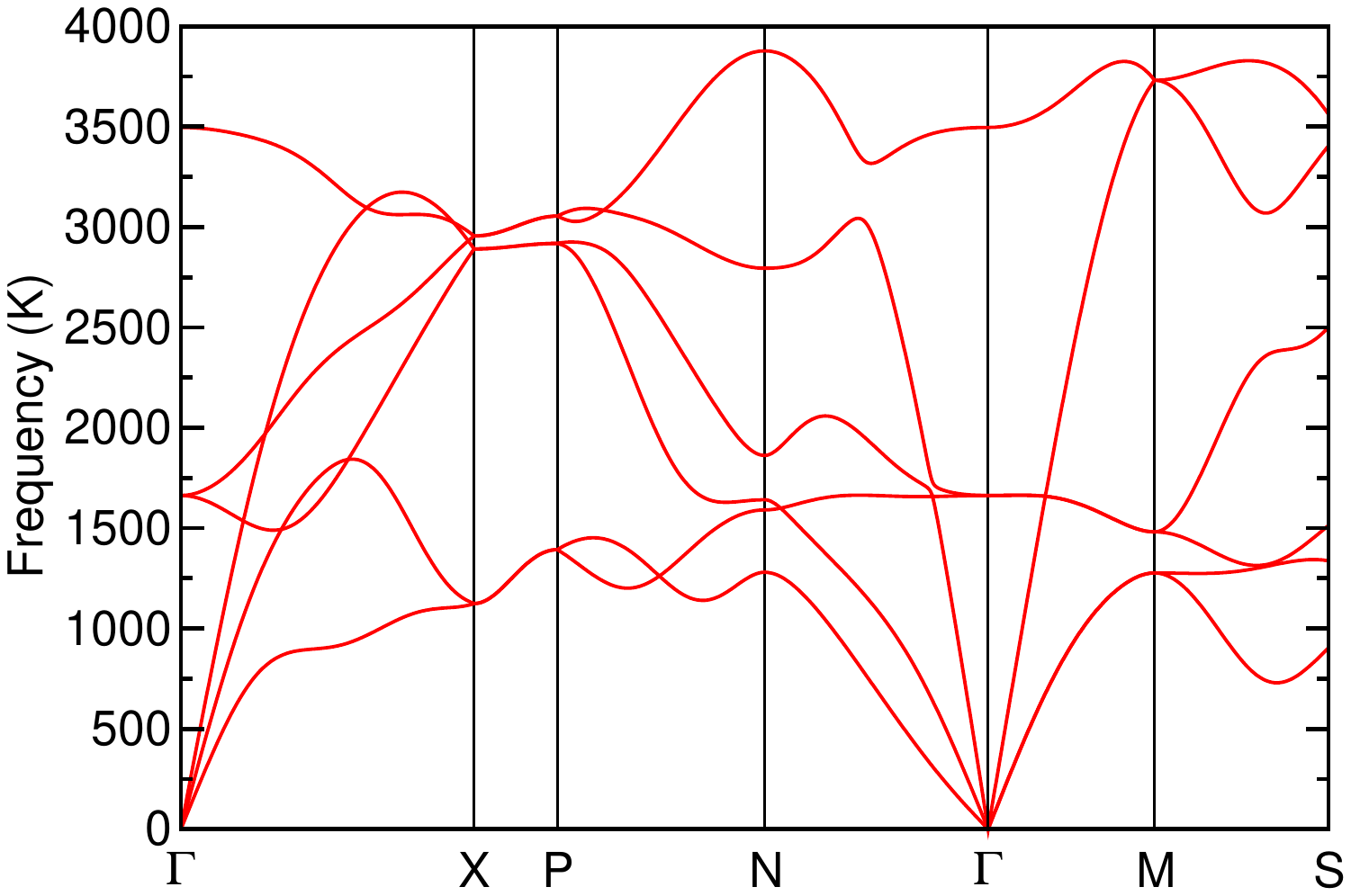}}}
\caption{Phonon dispersion of atomic hydrogen calculated at $400$ \,GPa.}
\label{dispcurves}
\end{figure}

\begin{figure}
{\scalebox{0.37}{\includegraphics{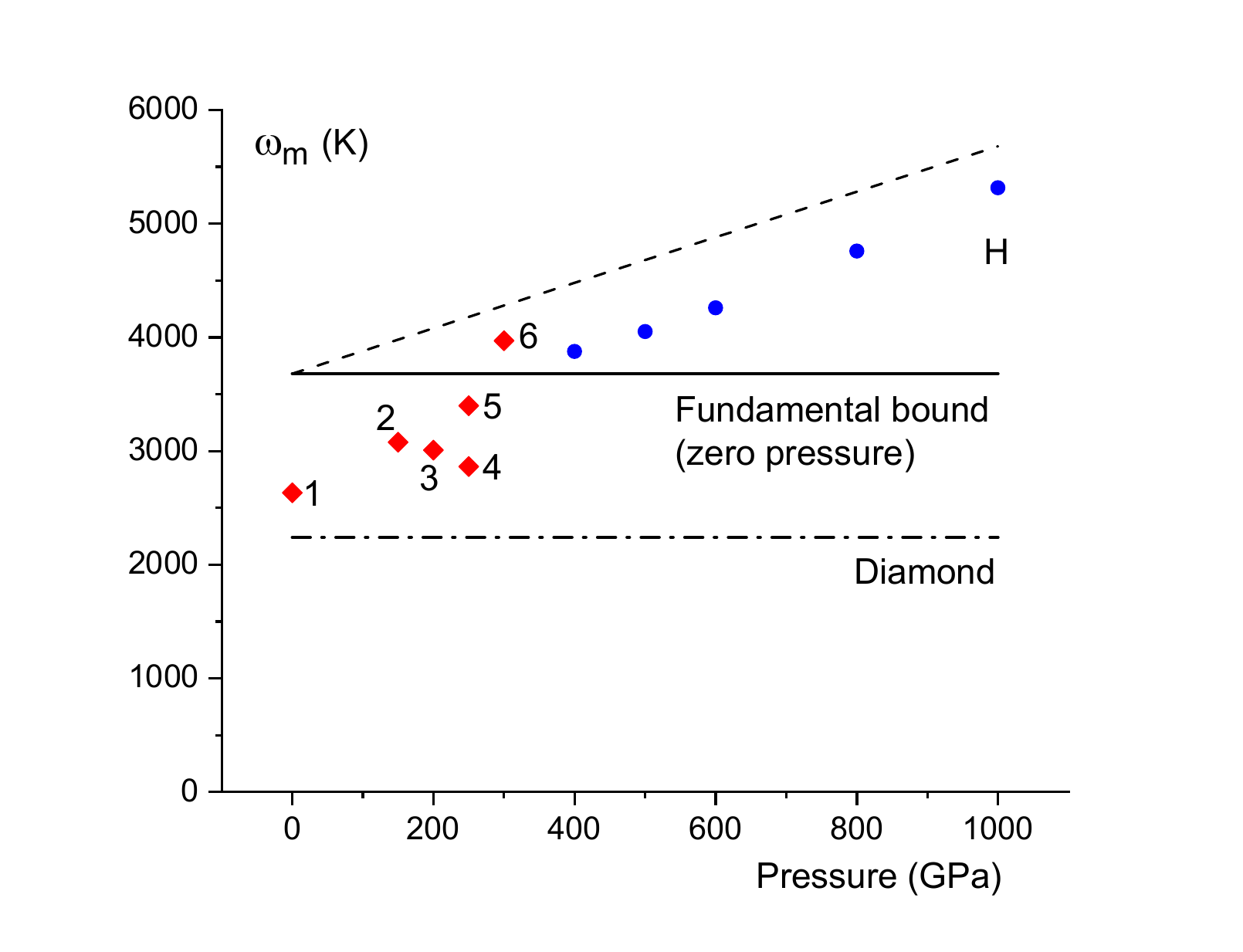}}}
\caption{Maximal phonon frequency as a function of pressure for different systems. Blue bullet points show the calculated maximal phonon frequencies in atomic hydrogen at the corresponding pressures. Red diamonds
show the maximal phonon frequencies in a range of hydrides at different pressures, which are, from left to right: MgIrH$_6$ (1) \cite{mgirh6}, MgVH$_6$ (2) \cite{mgvh6}, CaH$_6$ (3) \cite{cah6},
H$_3$S (4) \cite{h3s}, LaH$_{10}$ (5) \cite{lah10} and SrH$_{10}$ (6) \cite{srh10}. The horizontal solid black line shows the fundamental bound $\omega_{\rm m}^{\rm u}$ \eqref{omegal}. The horizontal dash-dotted line shows the Debye frequency of diamond at ambient pressure. The dashed line on top of the graph shows $\omega_{\rm m}^{\rm u}(P)$ in Eq.\,\eqref{externalp} with $a=2a_{\rm B}$.}
\label{bound}
\end{figure}

A representative phonon dispersion for atomic hydrogen at $400$\,GPa is shown Fig.\,\ref{dispcurves}, with the maximal phonon frequency occuring at the N point of the Brillouin zone with reciprocal
space coordinates $(0,\frac{1}{2},0)$. The maximal frequencies in the calculated phonon dispersion curves are shown in Fig.\,\ref{bound} as a function of pressure, and they are located at the N point
for pressures below about 600\,GPa and at the $\Gamma$ point for pressures above that value. The lowest pressure depicted corresponds to a pressure above the dynamical stability range of the atomic
hydrogen structure. The diagram in Fig.\,\ref{bound} also shows the upper bound \eqref{omegal} and $\omega_{\rm D}$ in diamond.

We observe in Fig. \ref{bound} that the calculated $\omega_{\rm m}$ is only slightly above $\omega_{\rm m}^{\rm u}$ at the high pressure of 400\,GPa in atomic hydrogen. Assuming a
pressure-independent effective $a$ in Eq. \eqref{externalp} for estimation purposes, we find that $\omega_{\rm m}^{\rm u}(P)$ lies close and above the hydrogen points if $a$ is taken to be about
$2a_{\rm B}$ (see dashed black line in Fig.\,\ref{bound}). This is consistent with the elementary volume being related to the Bohr radius and fundamental constants. Although $a=2a_{\rm B}$ in
Eq.\,\eqref{externalp} introduces an \textit{ad hoc} choice, the dashed line in Fig.\,\ref{bound} demonstrates that the pressure effect on $\omega_{\rm m}^{\rm u}$ can be accounted for with a
sensible and physically justified choice of $a$. In this regard, we note more generally that evaluating bounds on many important condensed matter properties using fundamental constants only
inevitably involves approximations \cite{advphysreview,brareview}. This includes $\omega_{\rm m}^{\rm u}$ in Eqs.\,\eqref{ratio2}-\eqref{omegal} which use approximate relations such as
Eq.\,\eqref{ratio}. Nevertheless, we see that the upper bound $\omega_{\rm m}^{\rm u}$ is in good agreement with computational results.

In addition to atomic hydrogen with the lightest atomic mass, another appropriate test for the theoretical upper bound $\omega_{\rm m}^{\rm u}$ in Eq.\,\eqref{ratio2} involves the comparison with
hydrogen-rich systems. These include hydride superconductors at high pressure, which have been of interest recently. These systems exhibit the highest superconducting critical temperature $T_c$
currently known extending to 250 K \cite{hreview1,hreview2,hreview3}. For our assessment of the maximal frequency upper bound, we chose a range of hydrides based on their chemical and structural
variety as well as a wide range of pressures at which phonons have been previously calculated using \textit{ab initio} methods. Our list includes data on LaH$_{10}$\,\cite{lah10} and
H$_3$S\,\cite{h3s}, with the highest experimentally measured $T_c$, two other binary hydrides CaH$_6$\,\cite{cah6} and SrH$_{10}$\,\cite{srh10}, and two ternaries MgIrH$_6$\,\cite{mgirh6} and
MgVH$_6$\,\cite{mgvh6}. The maximal phonon frequencies extracted from the calculated dispersion curves of these hydrides are shown in Fig.\,\ref{bound} as red diamonds. We observe that most points
are below the fundamental bound $\omega_{\rm m}^{\rm u}$. The maximal frequency in SrH$_{10}$ at the highest pressure of 300\,GPa exceeds the fundamental bound \eqref{omegal} but is below the bound
accounting for the pressure effect \eqref{externalp}.

Our derivation of the upper frequency bound is mostly suited to evaluate the highest oscillation frequency of acoustic phonon modes or in systems with no substantial hierarchy of bonding energies and
frequencies (recall that we apply Eq. \eqref{rydberg} for the bonding energies in the system). On the other hand, the intra-molecular frequencies in molecular solids (``vibrons'') can be considerably
higher than the rest of the frequency spectrum, and the highest oscillation frequency can be higher than $\omega_{\rm m}^{\rm u}$ in these systems. Considering, for example, H$_2$ molecule and using
the reduced mass $m/2$ increases $\omega_{\rm m}$ in Eq. \eqref{ratio} and $\omega_{\rm m}^{\rm u}$ in Eq. \eqref{omegal} by $\sqrt{2}$. Recalling that the approximate derivation of Eq. \eqref{ratio}
dropped $\sqrt{2}$ on the right-hand side, the highest oscillation frequency in hydrogen-containing molecules and molecular solids can increase by a factor of 2. Accordingly, one could choose to
define the upper bound for the oscillation frequency as $2\omega_{\rm m}^{\rm u}=7360$ K, in which case it would include molecules and molecular solids: highest intra-molecular frequencies are about
6200-6400 K and lower than $2\omega_{\rm m}^{\rm u}$. This could be of interest from the point of view of having a maximal oscillation frequency in all systems including molecular ones. However, we
note that in view of the approximations involved in Eqs. \eqref{ratio} and \eqref{ratio2}, one of the main purposes of Eq. \eqref{omegal} is to evaluate the order of magnitude of the maximal
frequency from fundamental constants only.

In addition to deriving the upper frequency bound, the two main results in this section are: (a) the upper bound applies well to non-molecular systems and (b) Eqs.\,\,\eqref{ratio2} and
\eqref{omegal} show that fundamental constants set $\omega_{\rm m}^{\rm u}$ on the order of $10^3$--$10^4$ K, the result we will use to discuss phonon-mediated superconductivity below.

\section{Superconducting temperature and search for room-temperature superconductivity}

\subsection{Upper limit to $T_c$}

We now discuss the implications of the upper bound of $\omega_{\rm m}$ for phonon-mediated superconductivity and the superconducting critical temperature $T_c$. Upper bounds of $T_c$ have been discussed on the basis of different mechanisms and have been related to various system properties, including the electron-phonon coupling constant $\lambda$, the energy distribution of valence electrons, the phonon and Fermi energies, instabilities involving phonons, electrons and other instability types, phase fluctuations, and stiffness (see, e.g., Refs. \cite{mcmillan,cohen,valence,varmaropp,esterlisnpj,esterlisprb,hazragraphene,chubukov,sadovskii-review,2407.12922} and references therein). More recent work has offered model examples where these bounds can be exceeded \cite{hofmannpj}.

Our results on the maximal phonon frequency suggest that the upper limit for $T_c$ in phonon-mediated superconductors may be governed at a deeper level by fundamental physical constants. We first consider the Migdal-Eliashberg (ME) theory of superconductivity involving the electron-phonon coupling relevant to high $T_c$ in hydrides. We will later discuss the alternative electron-electron coupling mechanism for superconductivity.

In the ME picture, $T_c=C\,\overline\omega$, where $C$ is a monotonically increasing function of $\lambda$ and $\overline\omega$ is an averaged phonon frequency \cite{cohen,carbotte}. More recent work explored effects beyond the ME theory and noted that they may be inessential for the superconducting state and $T_c$ \cite{chubukov}. There are several other interesting insights related to the effects of phonons and $\lambda$ on $T_c$. First, depending on the system, different phonon modes may contribute differently to $\lambda$ and $T_c$. This has been recently discussed in hydrides where different hydrogen-dominated modes combine with other element-dominated modes to set $\lambda$ \cite{lah10,h3s,srh10,cah6,mgvh6,mgirh6}. Second, a more general insight is that $T_c$ has a maximum at a certain $C$ due to the competition between the increase of $T_c$ with $\lambda$ and the suppression of $T_c$ due to instabilities at large $\lambda$ \cite{cohen,esterlisprb,esterlisnpj,carbotte,roadmap} (see Ref. \cite{zaccone-lambda} of the effect of soft phonons in this regard). Supported by model simulations, theory, and empirical data, $C$ is $C\approx 0.1$ \cite{esterlisprb,esterlisnpj} and increases to 0.2-0.3 for a different type of instability \cite{2407.12922}.  Assuming that different types of instabilities all operate, the upper bound to $T_c$ corresponds to the lower range of $C$ which induces the instability and re-organisation first, with the associated lowering of $T_c$. Interestingly, a similar $C$ in the range $0.1-0.2$ corresponds to $\lambda=1-2$ \cite{cohen,carbotte}, which are typical values of $\lambda$ in hydrides under pressure \cite{chrisprb} where the highest $T_c$ is observed \cite{hreview1,hreview2,hreview3}. We are interested in an upper bound for $T_c$, hence we substitute the average $\bar\omega$ by the maximal phonon frequency $\omega_{\rm m}$. Then, using Eq. \eqref{ratio2} for $\omega_{\rm m}^{\rm u}$ at zero pressure, the upper bound of $T_c$, which we call $T_c^{u}$, is

\begin{equation}
T_c^{u}=C\omega_{\rm m}^{\rm u}=C\left(\frac{m_e}{m_p}\right)^{\frac{1}{2}}\frac{1}{32\pi^2\epsilon_0^2}\frac{m_ee^4}{\hbar^3}
\label{tc}
\end{equation}

Using $\omega_{\rm m}^{\rm u}$ set by fundamental constants in Eq.\,\eqref{omegal} and $C$ in the above range $0.1-0.2$, Eq.\,\eqref{tc} gives $T_c^{u}$ in the range $370$--$740$\,K and on the order of

\begin{equation}
T_c^{u}=10^2-10^3\,\rm {K}
\label{tc1}
\end{equation}

The scale of $T_c^{u}$ in Eqs.\,\eqref{tc}-\eqref{tc1} is set by fundamental physical constants (the order of magnitude of $C$ is not variable). This has interesting implications to which we will return later.

The range \eqref{tc1} can be verified numerically from ME theory, where the central quantity is the phonon-frequency-resolved coupling strength, given by the Eliashberg function $\alpha^2F(\omega)$. Restricting $\alpha^2F(\omega)$ to frequencies below the upper bound $\omega_m^u$ \eqref{omegal}, we can directly calculate an upper bound on $T_c$ as

\begin{equation}
    T_c^{u} \sim \max_{\alpha^2F \in S} T_c[\alpha^2F]
\label{eq:alpha2f_optimization}
\end{equation}

\noindent where the functional $T_c[\alpha^2F]$ is the critical temperature from the solution of the Eliashberg equations \cite{marsiglio-eliashberg-theory} and $S$ is a set of sensible Eliashberg functions obeying the upper bound on $\omega$, with a $\lambda$ value limited to avoid instabilities as discussed earlier \cite{cohen,esterlisprb,esterlisnpj,carbotte,roadmap}. Details of this optimization procedure are available in the Appendix.

Such an optimization is shown in Fig.\ \ref{fig:optimal_a2f} for a fixed value of $\lambda = 2$, and converges to $T_c^{u} \sim 600$ K, well within the range of Eq. \eqref{tc1}. Since $T_c$ is the result of a trade-off between $\lambda$ and $\bar{\omega}$ \cite{zaccone-lambda} one would expect that, in the face of a $\lambda$ limited by instability, $\bar{\omega}$ becomes the central quantity. Indeed, we see that the optimization proceeds by shifting spectral weight to as high a frequency as possible, corroborating other work employing dispersionless (Einstein) phonons to obtain an upper bound to $T_c$ \cite{2407.12922}, and confirming that the value of $\omega_m^u$ is the limiting factor in the optimization.

\begin{figure}
    \centering
    \includegraphics[width=\columnwidth]{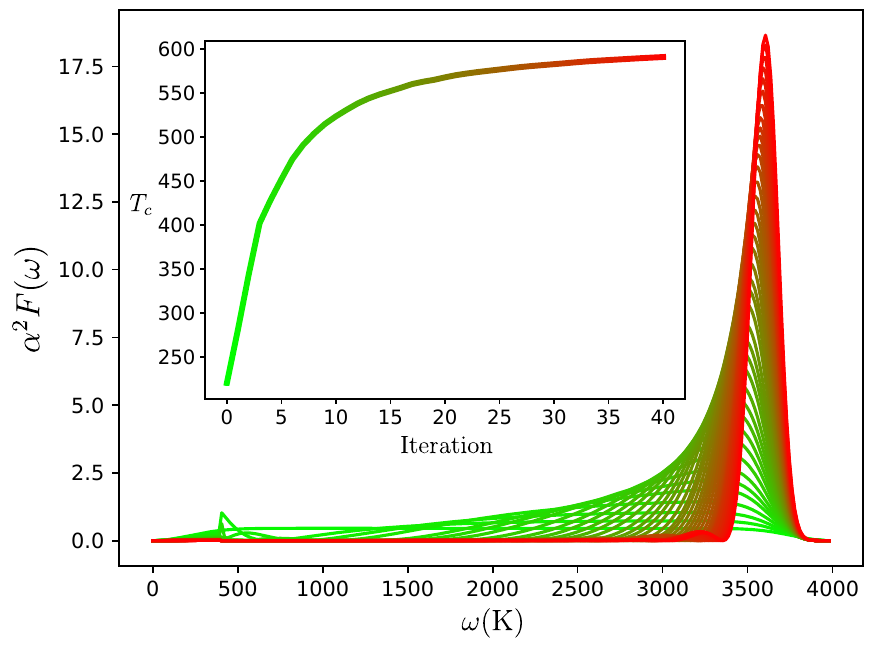}
    \caption{The evolution of the Eliashberg function $\alpha^2F(\omega)$ and corresponding $T_c$ (inset) as Eq. \eqref{eq:alpha2f_optimization} is optimized. Red (green) lines are later (earlier) iterations in the optimization.}
    \label{fig:optimal_a2f}
\end{figure}

$T_c^{u}$ in Eq.\,\eqref{tc} can be made more stringent and adaptable to a particular family of materials. Recall that (a) $E_{\rm R}$ used to evaluate the bonding energy $E$ is larger than bonding
energies in real materials and (b) the average mass in real materials is larger than the proton mass $m_p$ featuring in Eqs.\,\eqref{ratio2} and \eqref{tc}. Both effects reduce $T_c^u$. According to
Eqs.\,\eqref{ratio}, \eqref{ratio2}, and \eqref{tc}, the difference between $E$ and $E_{\rm R}$ results in $T_c^u$ decreasing by a factor $\frac{E_{\rm{eff}}}{E_{\rm R}}$, where $E_{\rm{eff}}$ is the
effective bonding energy in the relevant subsystem. The difference between the average mass and $m_p$ reduces $T_c^u$ by a factor $\left(\frac{m_{\rm{eff}}}{m_p}\right)^{\frac{1}{2}}$, where
$m_{\rm{eff}}$ is the effective mass in the relevant subsystem. The same approximate reasoning applies to the upper frequency bound \eqref{ratio2}.

We note that we do not consider the effect of pressure on $T_c$. $T_c$ can be a non-monotonic function of pressure and involve complex effects including, for example, effects of pressure on phonon frequencies, $\lambda$, screened Coulomb pseudopotential and so on \cite{pressure-tc}. In particular, to repeat the analysis in this work, the affect of pressure on the instability limit to $\lambda$ would need to be accounted for.

We also note that our bound \eqref{tc} invited a recent discussion in Ref. \cite{sadovskii1}. This work derived an upper bound to $T_c$ using a
different type of arguments and interestingly showed that this bound is expressed as the same combination of fundamental constants as Eq. \eqref{tc}.

\subsection{Other coupling mechanisms}

The above discussion also applies to systems where superconductivity is related to mechanisms other than the electron-phonon coupling. For electron-electron pairing, $T_c$ depends on the Fermi energy $E_{\rm F}$ and is on the order of about $\frac{E_{\rm F}}{10^2}$ \cite{varmaropp,uemura}. $E_{\rm F}\propto\frac{\hbar^2}{m_ea_{\rm B}^2}\propto\frac{m_ee^4}{\hbar^2}$ and depends on fundamental constants as $E_{\rm R}$ in Eq. \eqref{rydberg} does. Hence, $T_c$ in the electron-electron systems such as cuprates also depends on on fundamental constants as it is in the electron-phonon superconductors.

Noting that the upper limit to $E_{\rm F}$ is set by $E_{\rm R}$ in Eq.\,\eqref{rydberg} and recalling that $T_c$ in the electron-electron systems is on the order of $\frac{E_{\rm F}}{10^2}$ \cite{varmaropp,uemura}, we observe that the upper bound to $T_c$ in these systems is close to $T_c^{u}$ because $C\left(\frac{m_e}{m_p}\right)^{\frac{1}{2}}$ in Eq.\,\eqref{tc} with $C\approx 0.1$ is on the order of about $\frac{1}{10^2}$.

\subsection{Fundamental constants and search for room-temperature superconductivity}
\label{superc}

Having seen that the overall scale of $T_c^{u}$ is set by fundamental physical constants, we now ask the important question of what would be the effects of fundamental constants having different values. This question is intensely researched in high-energy physics and astrophysics and is related to fine-tuning of the Universe and associated grand challenges in modern science \cite{barrow,barrow1,carr,carrbook,finebook,cahnreview,hoganreview,adamsreview,uzanreview,grandest}. How would superconductivity change were fundamental constants (e.g., $\hbar$, $m_e$, $e$ and $\frac{m_e}{m_p}$ in Eq.\,\eqref{tc}) to take different values? This question is well-posed because it is possible to change $T_c^{u}$ in Eq.\,\eqref{tc} by varying $\hbar$, $m_e$, $e$ and $m_p$ while keeping $\beta=\frac{m_e}{m_p}$ and $\alpha=\frac{e^2}{\hbar c}$ unchanged and hence maintain the rest of essential processes (e.g. those involved in stellar formation and evolution, synthesis of heavy elements, stability of ordered structures and so on) intact \cite{barrow}.

We observe that were different fundamental constants to give $T_c^{u}$ on the order of, for example, $10^{-6}$ K or lower, superconductivity would be unobserved. Were different fundamental constants to give $T_c^{u}$ on the order $10^{6}$ K, many superconductors would have $T_c$ in excess of 300 K. Consistent with Eqs.\,\eqref{tc}-\eqref{tc1} giving $T_c^{u}$ on the order of $10^2-10^3$ K due to currently observed fundamental constants, we observe superconductivity in the temperature range $T\lesssim$ 100-200\,K. This understandably stimulates the current research into finding systems with $T_c$ of 300\,K and above.

Therefore, the very existence of the current line of enquiry to identify systems with $T_c$ above 300\,K is itself due to the values of fundamental constants currently observed.

\section{Fundamental constants and observability of phenomena in condensed matter physics}

The discussion in the previous section implies that fundamental constants can affect observability of an entire condensed matter {\it phenomenon} such as superconductivity. This implication is wider than what was discussed in earlier work
\cite{kss,sciadv,spin,hbarm1,hbarm2,hbarm3,hbarm4,hartnoll,hartnoll1,behnia,nussinov1,momentumprb,sciadv2,brareview,sciadv2023,advphysreview,myropp}
where fundamental constants were shown to set a bound on a {\it property} (e.g., viscosity, thermal conductivity, diffusivity, or speed of sound) which is {\it already} observed.

The observability of superconductivity as a result of current values of fundamental constants can be extended more generally to the observability of other phenomena and phase transitions. Indeed, different fundamental constants substantially increasing $\omega_{\rm m}^{\rm u}$ in Eq. \eqref{ratio2} would result in $T\ll\hbar\omega_{\rm m}$ and hence $T\ll\hbar\omega_{\rm D}$ ($\omega_{\rm D}$ is Debye frequency) at typical planetary temperatures. This would result in the solid properties being always quantum and the classical regime $T>\hbar\omega_{\rm D}$ unobservable. This would also suppress or arrest second-order structural phase transitions because of the smallness of the phonon entropy $S\propto\left(\frac{T}{\hbar\omega_{\rm D}}\right)^3$ when $T\ll\hbar\omega_{\rm D}$ \cite{landaustat}.

Another example is the first-order melting transition where melting temperature and characteristic slopes of the pressure-temperature melting lines are set by fundamental constants \cite{premelting}. This implies that different values of these constants would result in melting either (a) always taking place at typical planetary conditions and no solids existing or (b) being unobservable (in this case, the ice melting temperature would be above typical planetary temperatures implying no water-based life).

Linking the discussion in the previous paragraph and in Section \ref{superc}, we note that current fundamental constants enable diverse phenomena such as water-based life and superconductivity.

This and similar future discussion deepen our understanding of how fundamental constants affect observability of entire new phenomena at condensed matter length and energy scales. This helps to fill the large gap \cite{myropp} between the scales involved in nuclear physics and astrophysics where fundamental constants were shown to give rise to different key effects such as the stability of nuclei, star formation, and heavy nuclei synthesis
\cite{barrow,barrow1,carr,carrbook,finebook,cahnreview,hoganreview,adamsreview,uzanreview}. Observability of each effect and phenomenon imposes constraints on fundamental constants \cite{barrow}. Discussing different constraints along a continuous path of length and energy scales, including condensed matter scales, is likely to bring about new insights into understanding fundamental constants.

There may be a sense in which bounds to different properties and observability of phenomena can be expected to be related to fundamental constants on general grounds. We add two remarks in this regard. First, it is important to show how and why this happens. Indeed, one view holds that the essence of a physical theory is to provide a relationship between different physical properties and experimental outcomes \cite{landaupeierls}. Our current discussion provides such relationships, including the relationship between bounds on vibration frequency, $T_c$, observability of phase transitions on one hand and fundamental constants on the other hand. Second, it is not clear apriori whether bounds can be expressed solely in terms of fundamental constants:
recently discussed bounds depend on the combination of fundamental constants and external parameters such as temperature \cite{spin,hbarm3,hbarm4,hartnoll,hartnoll1,behnia,nussinov1}. On the other hand, our bound on oscillation frequency \eqref{ratio2} solely depends on fundamental constants. This is a more general and interesting result. This is particularly so in view that phenomena such as superconductivity, phase transitions and so on involve many-body collective effects which are considered as emergent and not reducible to fundamental constants (see Introduction).

\section{Summary}

In summary, we have shown that fundamental physical constants set the upper limit to the vibrational frequency of atoms in condensed (non-molecular) systems. This bound is in agreement with
simulations of atomic hydrogen and high-temperature superconducting hydrides. We have also proposed that fundamental constants set the upper limit to the superconducting critical temperature on the
order of 10$^2$--$10^3$\,K. This implies that the current line of enquiry to discover superconductors above 300\,K is itself due to the observed values of fundamental constants. This has led us to
observing that fundamental constants affect the observability of entire new effects and phenomena in condensed matter physics.

\begin{acknowledgments}
B.M. acknowledges funding from a UKRI Future Leaders Fellowship [MR/V023926/1], from the Gianna Angelopoulos Programme for Science, Technology, and Innovation, and from the Winton Programme for the
Physics of Sustainability. The computational resources were provided by the Cambridge Tier-2 system operated by the University of Cambridge Research Computing Service and funded by EPSRC
[EP/P020259/1]. K.T. is grateful to V. Brazhkin for discussions and EPSRC for support.
\end{acknowledgments}

\appendix

\section*{Appendix. Optimization of the Eliashberg function}

In isotropic Migdal-Eliashberg theory, $\lambda$ is a functional of the Eliashberg function $\alpha^2F(\omega)$, specifically
\begin{equation}
    \lambda[\alpha^2F] = 2\int \frac{\alpha^2F(\omega)}{\omega} d\omega.
\label{eq:lambda_functional}
\end{equation}
The critical temperature is obtained by solution of the Eliashberg equations \cite{marsiglio-eliashberg-theory}, whose only inputs are the Eliashberg function and a value for the Coulomb pseudopotential $\mu^*$. We can therefore write $T_c = T_c[\alpha^2F](\mu^*)$, a functional of $\alpha^2F$ and a function of $\mu^*$.

We can obtain an upper bound for the critical temperature by extremizing $T_c[\alpha^2F(\omega)](\mu^*)$ over a set of sensible $\alpha^2F$ functions, for a typical value of $\mu^*=0.125$. We note that $\alpha^2F(\omega)$ can not extend beyond the maximum phonon frequency in the system: $\alpha^2F(\omega > \omega_m) = 0$. For a finite $\lambda$ value, $\alpha^2F(\omega)$ must also smoothly approach $0$ as $\omega \rightarrow 0$. Finally, $\alpha^2F(\omega)$ must be positive. These form our conditions for a sensible Eliashberg function. As noted in the main text, $\lambda$ is naturally limited to ${\lambda \sim 1 - 2}$ by instabilities. To calculate an upper bound, we therefore fix $\lambda = 2$. For the function space $F$ containing $\alpha^2F$, we can construct a map $M : F \rightarrow F$ whose image space satisfies these constraints on $\alpha^2F$. That is to say, $\forall f \in F$,
\begin{align}
    \lambda[M(f)] &= 2 \label{eq:lambda_condition},\\
    M(f)(\omega) &\geq 0 \;\forall \omega, \label{eq:positivity_condition} \\
    \lim_{\omega \rightarrow \omega_m} M(f)(\omega) &= 0, \label{eq:zero_freq_condition}\\
    \lim_{\omega \rightarrow 0} M(f)(\omega) &= 0. \label{eq:max_freq_condition}
\end{align}
In particular, noting that Eq. \eqref{eq:lambda_functional} is linear in $\alpha^2F$, we can satisfy Eq. \eqref{eq:lambda_condition} by taking
\begin{equation}
    M(f) = \frac{2E(f)}{\lambda[E(f)]}
\end{equation}
For some $E: F \rightarrow F$. We choose
\begin{equation}
    E(f)(\omega) = f^2(\omega) G(\omega) G(\omega_m - \omega)
\end{equation}
where mapping $f \rightarrow f^2$ ensures positivity (Eq. \eqref{eq:positivity_condition}) and $G$ is an envelope function, which we take as
\begin{equation}
    G(\omega) = \begin{cases}
        1 - \exp{(-\omega^2/\sigma^2)} \text{ if } \omega > 0 \\
        0 \text{ otherwise}
    \end{cases}
\end{equation}
so that the limits in Eqs. \eqref{eq:zero_freq_condition} and \eqref{eq:max_freq_condition} hold, and are approached smoothly. We take the width of the envelope to be $\sigma = 300$ K, but this value makes little difference to the resulting $T_c$.

Having constructed $M$, we can perform the constrained optimization of $T_c$ as
\begin{equation}
    T_c^{u} = \max_{f \in F} T_c[M(f)](\mu^*=0.125)
\end{equation}
by representing $f$ on a uniform grid of frequency points and employing the BFGS algorithm \cite{fletcher_1987}. We begin the optimization with an initial guess of $f = 1$.

A code to perform this optimization is available \cite{optimala2f}.


\end{document}